# The Case for a General and Interaction-based Third-party Cookie Policy

Istemi Ekin Akkus
MPI-SWS
iakkus@mpi-sws.org

Nicholas Weaver
ICSI & UC Berkeley
nweaver@icsi.berkeley.edu

*Abstract*—The privacy implications of third-party tracking is a well-studied problem. Recent research has shown that besides data aggregators and behavioral advertisers, online social networks also act as trackers via social widgets. Existing cookie policies are not enough to solve these problems, pushing users to employ blacklist-based browser extensions to prevent such tracking. Unfortunately, such approaches require maintaining and distributing blacklists, which are often too general and adversely affect non-tracking services for advertisements and analytics. In this paper, we propose and advocate for a general third-party cookie policy that prevents third-party tracking with cookies and preserves the functionality of social widgets without requiring a blacklist and adversely affecting non-tracking services. We implemented a proof-of-concept of our policy as browser extensions for Mozilla Firefox and Google Chrome. To date, our extensions have been downloaded about 11.8K times and have over 2.8K daily users combined.

## I. INTRODUCTION

To obtain extended web analytics (e.g., user demographics, other sites visited), web publishers often outsource the collection of analytics information to third-party data aggregators. These aggregators may also provide behavioral advertisements that are tailored to a user's interests based on his/her browsing history. Similarly, publishers also embed social widgets provided by online social network (OSN) providers (e.g., Facebook Like) on their sites to increase user engagement and the site's exposure to other users. Aggregators and OSN providers also benefit from this setup: they obtain vast amounts of information about users' browsing behavior across the web and use this information via targeted advertisements [3].

Unfortunately, these actions come with a price for user privacy and raise concerns about users being tracked on the web. This tracking enables these third parties to compile detailed behavior of individual users with the sensitive information they obtain [32]. Thus, these parties are given a lot of information about users' actions on the web and have to be trusted not to abuse it. This trust has been violated in the past [6, 16, 17].

Existing cookie policies in modern browsers offer a partial solution to this problem. A user can select a cookie policy, in which the browser will not allow any third-party cookies to be set. While effective at preventing tracking for behavioral advertisements, this policy causes issues when the third party is an OSN provider. For example, Firefox's "Never accept third party cookies" policy breaks the functionality of social widgets on publisher sites and does not allow a user to interact with them, even when the user is logged in to the OSN that provided the widgets. Chrome and Safari behave the same way.

Another privacy option in Firefox aims to solve exactly this problem, such that cookies from third parties will be accepted if the user has visited the third party site as a first party in the past.[1] For example, if the user is logged in to Facebook, all Facebook Like buttons on other publishers will function properly. This option, however, allows the OSN provider to act as a tracker and learn about the user's visit to the publisher, even if the user did not interact with the social widget [25, 32]. Tools, such as Priv3 [12, 25] and ShareMeNot [19, 32], aim to prevent this tracking by OSN providers.

Following these tools, other popular client-side privacy tools like Ghostery [4] and Disconnect [2] started preventing social widgets from being loaded, in addition to the trackers by data aggregators. Using a blacklist of aggregators, behavioral advertisers and OSN providers, these tools scan the loading page and prevent blacklisted elements from being loaded.

This blacklist-based approach has several limitations. First of all, the blacklist needs to be maintained and then distributed to clients in a timely fashion; otherwise, the benefits of using such a tool are greatly reduced. These tools try to find an optimum update schedule for their tracker libraries. For example, Ghostery regularly updates its library of trackers while Disconnect checks for updates every day. Other less popular tools like Priv3 and ShareMeNot support only a handful of third parties, and depend on their developers to keep up with new social widgets. This maintenance of the blacklist can be cumbersome and error-prone: there is no guarantee that *all* third-party trackers will *always* be included in the blacklist.

In addition, such a blacklist may be bypassed by third parties, simply with a configuration trick at their servers. For example, Apache2 directive 'AliasMatch' [9] enables a third party to serve the blacklisted element (i.e., JavaScript file, social widget) via customizable URLs, such that each publisher uses a different source, yet the third party serves the same file.[2] This trick would force the tools to blacklist entire domains, which can become problematic if legitimate files not related to tracking are also served (e.g., libraries, images, OSN site).

Finally, these blacklists are very broad: They include first-party analytics tools, such as Piwik, Open Web Analytics and Mint Analytics [8, 10, 11], preventing publishers from learning about their users' behavior on their own sites. They also include non-behavioral advertising that, by definition, does not require the tracking of users across the web (e.g., Project

---
[1]Similar to Safari's "block cookies from third parties and advertisers".
[2]Demo available at https://nta.mpi-sws.org/test2/test.html

Wonderful). As a result, publishers who choose more privacy-friendly solutions for analytics and advertisements by *not* using third-party tracking are being unnecessarily penalized. While such non-tracking services can be removed from the blacklist, the maintenance issue is only amplified: the tool provider now has to categorize and determine which solutions are acceptable.

Privacy Badger from EFF [15] and Lightbeam from Firefox [7] dynamically track the third parties the browser is contacting. Privacy Badger aims to detect and block third parties that appear to be tracking the user without his/her consent. If the content is deemed necessary for the page, cookies are removed from these requests. Privacy Badger uses the same blacklist as ShareMeNot for OSN sites, and the interaction with the widget is blocked if not manually overridden. Lightbeam visualizes the third parties, which the user can block manually, but such blocking breaks the functionality of the widgets.

In this paper, we propose and advocate that modern browsers should implement a general cookie policy, such that third-party cookies will only be sent when and if the user interacts with the third party content. This policy supports social networking features when the user wants them, striking a balance between user engagement for publishers and privacy for users. This policy is general, because there is no blacklist to maintain, and thus, it is free of the above problems. Finally, this policy is also effective in preventing third-party tracking via cookies without punishing web analytics and advertising platforms that do not track users across the web.

The next section presents our goals. We list our assumptions in Section III. Section IV explains our design and how it achieves our goals. We describe our implementation in Section V, and discuss our policy's implications in Section VI. We conclude with future work in Section VII.

## II. GOALS

We advocate that the users should be the final decision makers with regards to tracking by third parties, be they data aggregators or OSNs. Specifically, we would like our policy to enable interactive features (i.e., social widgets) in an on-demand fashion. Like previous approaches [12, 25, 32], we think that user interaction is necessary to achieve balance for privacy and functionality for a social web.

At the same time, we would like to devise a general cookie policy for third parties to prevent third-party tracking by not only OSNs but also data aggregators and advertisers. This policy should not depend on a blacklist unlike the above tools; thus, it should not require the cooperation of a developer to maintain and distribute such a blacklist to protect user privacy.

Finally, our policy should not interfere with non-tracking services for analytics and advertisements, and penalize publishers using such arguably more privacy-friendly services. We recognize the fact that for a sustainable web, publishers need statistical data about their users to improve their services as well as advertisements to financially support their operations.

## III. ASSUMPTIONS

We consider any content that is not loaded from the first party domain as third party content. Such content can include social widgets from OSN providers as well as advertisements.

We assume that the cookie preferences reflect the users' intentions and that the third parties are not going to try to bypass them. In a recent example, Doubleclick was caught deliberately circumventing Safari's default policy, and got sued by the Federal Trade Commission [5]. We assume such attempts are frowned upon, if not illegal, and the attempting party risks its reputation. We think that the data aggregators providing voluntary opt-out mechanisms already show their good faith in this regard, and that this assumption is reasonable.

More specifically, we leave methods to circumvent user cookie preferences outside our scope. One such method is fingerprinting, in which a third party creates a unique signature of a user's browser by combining various pieces of information in the browser environment, such as plugins, fonts and resolution. This fingerprint is then used to track the user across websites [26, 28] without storing any cookies on the user's device. With the prevalence of such practices increasing [21, 22, 30], potential defenses are already being researched [29].

Another method outside our scope is 'cookie synching' [1, 20]. In cookie synching, publishers share first-party cookie values of their users with third parties, by embedding a resource request to the third party with the first-party cookie values as parameters, enabling the third parties to set their own cookies. As a result, they can establish a mapping between the received cookie values and the cookie values they set, such that the user's browsing behavior can be correlated.

Finally, we assume that mashups, sites with data and code from multiple publishers, are interactive. If not, we assume that they will continue to function without the third parties receiving user-specific data (i.e., cookies). For example, most mashups using the Google Maps API still function if the user is not logged in to Google.

## IV. DESIGN

To achieve our goals, we propose the following policy: Any content from a third party domain (e.g., social widgets, advertisements) should be loaded without the associated (third-party) cookies. This content will be reloaded with the associated cookies, when (and if) the user interacts with it.

**Reload-on-click:** We define user interaction as the mouse click to a page element. Although other events such as hovering over an element or key presses can also constitute user interaction, we think that a click covers a big portion of user interactions with content. Other approaches also use a click as an indication of user intent to interact with social widgets [4, 12, 19].

Our previous work with other collaborators, Priv3, showed that reloading social widgets after the user click is effective for enabling social features on a website without compromising user privacy for functionality [12, 25]. For example, when the user wants to click the Facebook Like button on a page, it is reloaded by sending the user's Facebook cookies. Our current work enhances this approach with two new mechanisms.

**Two-click control:** After the third-party content is initially loaded without sending the user's cookies, we use a two-click control. The first click enables the third-party content by reloading the third-party content with the user's cookies. The second click registers the original action. For example, when the Facebook Like button is reloaded after the (first) user click, it shows information about the friends of the user who also liked that page. If the user wants to like the page, the second click will register the action. In this case, Facebook knows about the user's visit to that page after the first click (i.e., activation of the widget). Enabling widgets in this manner still provides functionality, but at the same time, allows OSN users not interested in using the social widgets to have more control over when OSNs can learn about their browsing.

**Generalization:** In contrast to previous approaches, our policy does not require a blacklist: it is applied to any third party content. Detecting such content is a straightforward task similar to the same origin policy already employed by browsers (§V).

Besides social widgets, our policy is also effective in preventing other third party tracking via cookies: the user cannot interact with 'invisible' elements (e.g., pixel tags, invisible iframes) that are used for behavioral advertising and data aggregation purposes. As a result, cookies associated with these third parties will never be sent, preventing them from tracking the user across the web. These elements do not need to be detected at runtime or enumerated in advance as in a blacklist, because the policy applies to any third-party content.

Finally, our policy does not interfere with first-party analytics tools, because these tools use cookies that belong to the first-party domain whose requests are not modified. Similarly, this policy does not interfere with non-tracking (i.e., non-behavioral) advertisement systems, which by definition do not use any tracking cookies to load advertisements.

## V. Implementation

We implemented a proof-of-concept of our proposed policy as browser extensions. Our implementation, Priv3+, is available for Firefox [13] and Chrome [14]. To date, it has been downloaded about 11.8K times and has over 2.8K daily users.

**Detecting third-party requests and removing cookies:** When a page is being loaded, our tool intercepts HTTP GET requests for the resources embedded in the page. Requests to the first-party domain are let through unchanged with their cookies. Requests for a resource from a third-party domain are only let through after removing the cookie values. These cookies are still present in the browser (i.e., they are not deleted), and used when the user interacts with the third-party content.

**Third-party cookie access:** Strictly speaking, our policy does not prevent third parties from setting cookies on the user's browser. One caveat of this approach is that they can receive these cookies later, if the user visits the third party site as a first party. This issue opens the possibility of the third party accumulating the browsing history of the user by setting its cookies and hoping the user visits its website. Although the probability of a user visiting a tracker's website may be low, this issue becomes more important if the third party is an OSN.

As in the original Priv3 implementation [12, 25], Priv3+ prevents third-party scripts from accessing cookie values until the user interaction. As a result, third parties cannot use scripts to compile a list of visited pages in the cookie values to receive them later. We think that refusing new third-party cookies can also solve this problem, but have not implemented this feature.

**Social widgets versus advertisements:** The lack of the blacklist forced us to develop a heuristic to distinguish between advertisement networks (where a click needs to be passed unchanged) and social widgets (which require a reload). The social widgets in a publisher website are usually loaded in a sandboxed container such as an iframe to prevent security issues. For example, social widgets from Facebook, Google and Twitter as well as commenting platforms, such as Disqus, are loaded in a single iframe. The user click simply triggers the reload of this iframe. This approach is different than other client-side tools, such as Ghostery, which requires the user to reload the entire page rather than a single element.

Similar to social widgets, third-party advertisements are also loaded within iframes. For clicked advertisements, the click is processed without reloading the container. During user interaction, the two cases are distinguished using a heuristic: an advertisement is usually present in nested iframes due to the real-time ad auction process and multiple redirections, whereas a social widget is contained within only one iframe.

**Limitations:** Our biggest limitation is that our heuristic may fail to distinguish a social widget and an advertisement loaded in a single (i.e., non-nested) iframe. A user click on the advertisement will trigger a reload, which may have an adverse effect such as creating an extra impression that otherwise would not have occurred. More importantly, the click on the advertisement may not register creating an undesired behavior. We plan to further investigate how prevalent this issue is.

**Preliminary performance evaluation:** We created user profiles in Firefox for each existing cookie policy as well as Priv3+. We visited up to 10 random pages from each of the top 1K sites from a snapshot of Quantcast top sites [18] from November 2014 for a total of 7,257 pages and recorded the load times. On average, Priv3+ adds 3.94% overhead compared with the default policy of accepting all third party cookies. Not accepting third party cookies and accepting third party cookies from visited sites add 1.73% and 1.32% overhead, respectively.

**Miscellaneous:** Priv3+ shows information about the third parties present in a web page. It can highlight different types of third-party content (e.g., content that had cookies, content that had no cookies). The user can create a whitelist of third parties to be loaded with cookies and on which publishers.

## VI. Discussion

**Reloading advertisement iframes with cookies:** One can argue that a clicked advertisement shows the user's intention to interact with the advertiser, and thus, the advertisement iframe should be reloaded. This approach has the following problems: First, it is not clear which iframe to reload because there may be multiple, nested iframes. Reloading the parent may generate a different child iframe containing another advertisement less relevant to the user. Some iframes do not have their 'src' property set preventing the reload. Most importantly, reloading may have adverse effects for the advertisers, triggering new auctions and double-charging for impressions and clicks. For these reasons, we pass the click to an advertisement unchanged.

**Evercookies:** Evercookies use storage vectors (e.g., Flash cookie store) [21], which are not deleted when the browser cookies are cleared. Trackers exploit these storage vectors to respawn old cookie values to achieve a longer persisting tracking period. Our policy would prevent these respawned cookies to be sent to third parties without user interaction.

**Cookie synching:** Our policy partially prevents cookie synching that uses previously set (third-party) cookies. The cookie access mechanism may prevent other methods like using first-party cookie values as parameters for third-party resources.

**Behavioral advertisements & extended web analytics:** Our policy prevents third-party tracking used for behavioral advertisements, which may be deemed necessary for a sustainable web. A byproduct of this tracking is extended web analytics, in which the aggregators can provide visitor demographics. There have been multiple efforts to provide behavioral advertising that is privacy-preserving and comparable to today's systems [27, 31, 33]. Similarly, previous research shows how the same aggregate information can be obtained without violating user privacy [23, 24]. We think these efforts as well as our policy are steps in the right direction to provide essential services for a sustainable web without compromising user privacy.

## VII. Conclusion & Future Work

We proposed and advocated a general, interaction-based third-party cookie policy. With our policy, third party content is loaded without sending associated third-party cookies, effectively preventing tracking by OSNs, data aggregators and behavioral advertisers. This policy strikes a balance between functionality of social networking and privacy by requiring user interaction to reload the social widgets with cookies when the user wants. Our policy is general and does not depend on a blacklist, automatically solving problems associated with maintenance, distribution and circumvention of the blacklist. Finally, it supports non-tracking analytics and advertisement services and does not penalize publishers who use these more privacy-friendly tools. Ideally, we would like our policy implemented and supported in major browsers.

Future work includes studying if our policy causes any adverse effects. First, we will expand our performance evaluation with more sites and compare our page load times with blacklist-based tools, which can prevent loading a blacklisted element and save time. Second, we will devise a methodology to quantify our policy's effects on rendered pages and determine if these effects lead to functionality issues. Advertisements, customization and changing content can cause differences making this task complex. We plan to compensate for these differences by comparing the structure of the pages (i.e., DOM tree). Finally, we hope to gather more users and obtain their feedback, which will help us better understand how users perceive and treat different types of third-party content.

## Acknowledgment

We thank our reviewers for their valuable feedback.